\documentclass[sigconf]{acmart}
\usepackage{amsmath}
\usepackage{mathtools}
\usepackage{amsthm}
\usepackage{array}
\usepackage{multirow}
\usepackage{tabularx}
\usepackage[table]{xcolor}
\usepackage{dsfont}
\usepackage{bbding}

\usepackage[capitalize,noabbrev]{cleveref}

\definecolor{aliceblue}{rgb}{0.94, 0.97, 1.0}
\definecolor{softblue}{RGB}{218, 232, 252}

\newcolumntype{x}[1]{>{\centering\arraybackslash}p{#1pt}}
\newcolumntype{y}[1]{>{\raggedright\arraybackslash}p{#1pt}}
\newcolumntype{z}[1]{>{\raggedleft\arraybackslash}p{#1pt}}

\AtBeginDocument{%
  }

\setcopyright{acmlicensed}
\copyrightyear{2026}
\acmYear{2026}
\acmConference[ACMMM '26]{Proceedings of the 34th ACM International Conference on Multimedia}{November 10--14, 2026}{Rio de Janeiro, Brazil}
\begin{document}

\title{Group-of-Latents: Perceptual Video Compression at Extreme Bitrates via Masked Latent Generative Modeling}


% Shaokang Wang, Jinchang Xu, Peidong Jia, Zhijian Hao, Siyuan Qian, Fei Zhao, Rui Ma, Xiaozhu Ju, Jian Tang, Xiaodong Xie, Shanghang Zhang, Huizhu Jia

\author{
Shaokang Wang\textsuperscript{*1}
Jinchang Xu\textsuperscript{*$\dagger$1}
Peidong Jia\textsuperscript{1} 
Zhijian Hao\textsuperscript{2} 
Siyuan Qian\textsuperscript{1} 
Fei Zhao\textsuperscript{1} 
Rui Ma\textsuperscript{1} \\
Xiaozhu Ju\textsuperscript{3} 
Jian Tang\textsuperscript{3} 
Xiaodong Xie\textsuperscript{1}
Shanghang Zhang\textsuperscript{1}
Huizhu Jia\textsuperscript{1}
}

\thanks{* Equal contribution. $\dagger$ Project leader.}
 % \Envelope \ Corresponding author.

\affiliation{
    \institution{
    \textsuperscript{1}State Key Laboratory of Multimedia Information Processing,
    School of Computer Science, \\ Peking University, Beijing, China
    }
    \city{}
    \country{}
}
\affiliation{
    \institution{
    \textsuperscript{2}State Key Discipline Laboratory of Wide Band-Gap Semiconductor Technology,\\
    School of Microelectronics, Xidian University, Xi'an, China
    }
    \city{}
    \country{}
}
\affiliation{
    \institution{
    \textsuperscript{3}X-humanoid, Beijing, China
    }
    \city{}
    \country{}
}

\email{
skwang6272@stu.pku.edu.cn, jinchang\_xu@pku.edu.cn
}

\renewcommand{\shortauthors}{Wang et al.}

\begin{abstract}
Most existing video compression algorithms follow a paradigm of transformation and quantization, optimizing the trade-off between distortion and bitrate. 
However, extremely low-bitrate compression remains an underexplored frontier where perceptual quality optimization under severely constrained coding resources has not been adequately addressed. 
In this paper, we propose a unified generative framework that leverages pre-trained Diffusion Transformer (DiT) priors to achieve high perceptual quality at extremely low bitrates. 
We first introduce a flexible \textit{Group-of-Latents} (GoL) strategy within the latent space of a causal tokenizer, explicitly partitioning the latent stream into intra $I$-latents and inter $P$-latents. 
The Deep Compression Module (I-DCM) then encodes key $I$-latents to preserve perceptual anchors with minimal overhead. 
Building upon these anchors, the DiT-based Unified Latent Denoising Module (U-LDM) refines intra-frame textures and synthesizes $P$-latents from noise, reconstructing temporal dynamics at zero additional bitrate cost. 
Extensive experiments demonstrate that our method uniquely operates in the extreme-low bitrate regime (e.g., $< 0.005$ bpp), achieving state-of-the-art perceptual fidelity with rich spatial details and robust temporal consistency. The code will be made publicly available.
\end{abstract}

\ccsdesc[500]{Computing methodologies~Computer vision tasks}

\keywords{Video compression, perceptual compression, diffusion model}


\maketitle

% ==========================================================================
\section{Introduction}
\label{sec:intro}

\begin{figure}[t]
  % \vskip 0.2in
  \begin{center}
    \centerline{\includegraphics[width=\columnwidth]{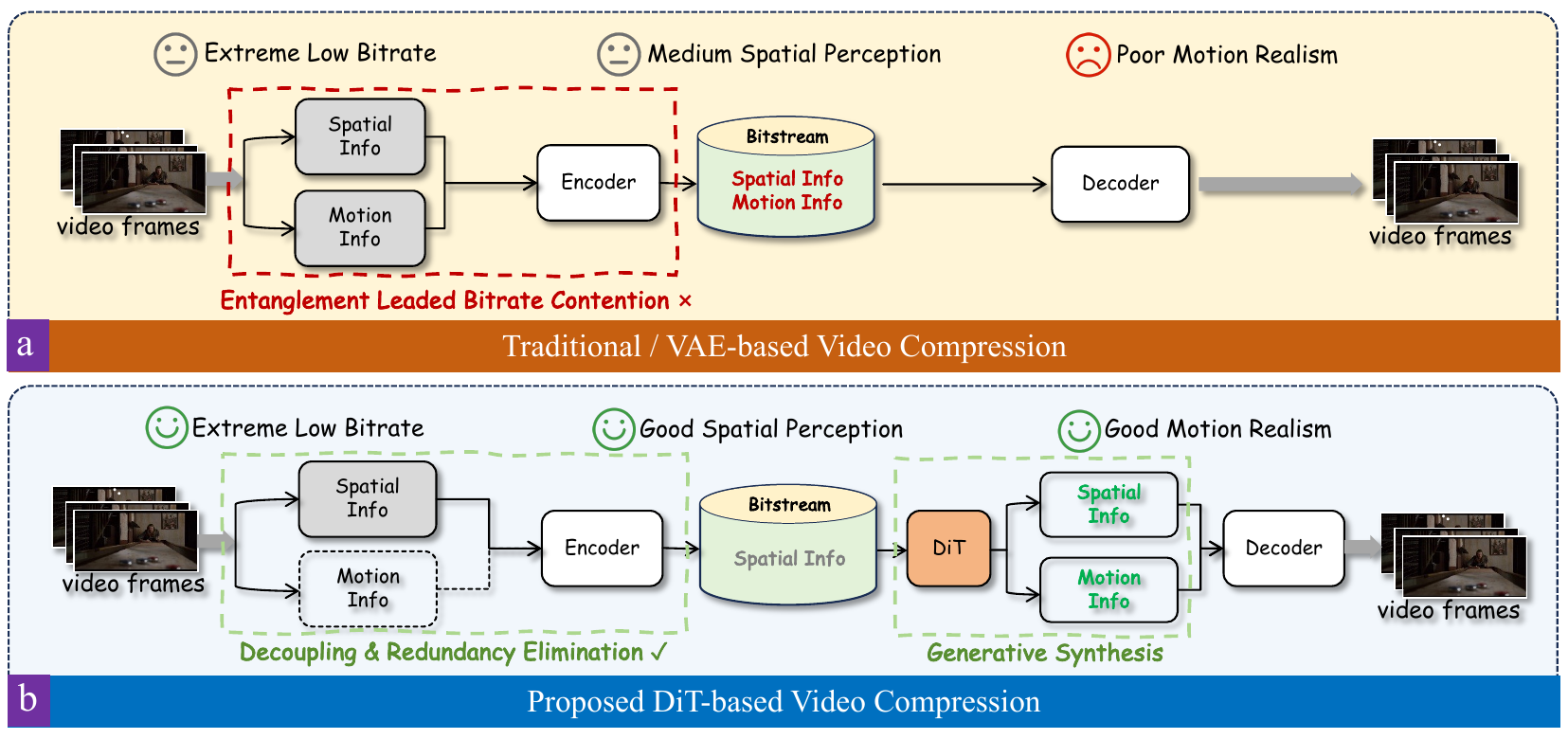}}
    \caption{Intuitive comparison of the proposed method with existing extreme-low bitrate compression algorithms. Existing methods (a) implicitly entangle texture and motion, leading to resource contention and inferior quality at low bitrates. In contrast, our design (b) explicitly decouples these tasks and leverages generative priors to independently ensure spatial perception and realistic motion.} 
    \label{fig:intuitive}
  \end{center}
  % \vskip -0.1in
\end{figure}

Amid the exponential growth of digital visual data, the development of high-efficiency video compression has become increasingly critical. Traditional codecs and Variational Autoencoder (VAE)-based end-to-end frameworks typically minimize distortion ($D$) (e.g., Peak Signal-to-Noise Ratio, PSNR) under rate constraints ($R_c$) through iterative refinement of handcrafted operators and neural architectures:
\begin{equation}
    \min D,\quad \text{s.t.}\ R \leq R_c.
\end{equation}
These established approaches have proven effective in achieving high-fidelity reconstruction within conventional bitrate regimes.

However, the exponential growth of visual data and the rise of bandwidth-critical applications~\cite{zhang2024cosmic,tao2024known} have rendered extremely low-bitrate compression a pressing necessity. More fundamentally, this regime raises a deeper question about the lower bounds of essential visual entropy required for intelligible reconstruction. Conventional codecs and VAE-based frameworks encounter a performance ceiling in this regime, where minimizing pixel-wise distortion leads to catastrophic perceptual degradation.
Rate-Distortion-Perception (RDP) theory~\cite{blau2018perception,blau2019rethinking} formalizes this limitation. Under extreme rate constraints, an inherent bottleneck on distortion minimization emerges, necessitating a pivot toward optimizing perceptual quality. RDP theory draws a fundamental distinction between sample-level signal fidelity and distribution-level perceptual realism. Traditional codecs, bound by pixel-wise distortion objectives, inevitably produce pervasive blur at extreme-low bitrates. Generative approaches, by contrast, prioritize perceptual alignment, achieving superior performance on both sample-level perceptual metrics (e.g., LPIPS) and distribution-level realism measures (e.g., KID).
This theoretical foundation motivates our focus on perceptual compression, synthesizing plausible visual details that transcend the artifacts inherent to distortion-optimized methods.

The success of generative models has catalyzed a paradigm shift in perceptual image compression~\cite{rombach2022high,gu2022vector,dhariwal2021diffusion,li2024blip,zhao2024uni,gupta2024photorealistic,wang2025lavie,zhou2024storydiffusion,jiang2024videobooth}, with GAN-based~\cite{agustsson2019generative,mentzer2020high,liu2021content,korber2025egic} and Diffusion-based~\cite{li2025toward,careil2024towards,zhang2025stablecodec,xu2025decouple} approaches achieving high-fidelity reconstruction at extreme bitrates. However, extending these gains to video remains a non-trivial challenge. Although initial attempts at extremely low-bitrate video compression have emerged~\cite{qi2025generative,wang2025tgvc,wang2025low}, they often yield suboptimal results.
We attribute this to the inherent complexity of video perception, which encompasses two coupled yet distinct dimensions: spatial texture and temporal motion. As illustrated in Figure~\ref{fig:intuitive}, existing models typically attempt to optimize both dimensions concurrently within VAE architectures that lack the capacity to disentangle them effectively. Under extreme bitrate constraints, the scarcity of encoding resources forces a compromise that satisfies neither dimension, resulting in degraded reconstructions. This instability fundamentally stems from treating video as a sequence of independent frames rather than a structured temporal entity.

To jointly optimize spatial and temporal perception at extremely low bitrates, we propose a generative video compression framework built upon pre-trained Diffusion Transformers (DiT). Recognizing that latent representations produced by causal tokenizers inherently preserve both intra- and inter-frame characteristics, we introduce the Group-of-Latents (GoL) structure, a latent-domain evolution of the traditional Group-of-Pictures (GoP) paradigm that strategically decouples the video into $I$-latent and $P$-latent components.
A Deep Compression Module (I-DCM) employs spatial downsampling and entropy modeling to encode $I$-latents into ultra-compact representations, establishing robust perceptual anchors at extreme bitrates. Since $I$-latents constitute only a fraction of the full latent sequence, a Unified Latent Denoising Module (U-LDM) leverages DiT priors to simultaneously refine quantization artifacts in $I$-latents and synthesize $P$-latents from pure noise. This mechanism reconstructs high-fidelity temporal dynamics at zero additional bitrate cost, effectively bypassing traditional residual coding.
The resulting $I/P$ decoupled design aligns with the fundamental structure of video perception and yields a versatile architecture. GoL configurations can be dynamically adjusted at inference time to navigate the trade-off between bitrate and perceptual fidelity without retraining.

In summary, our contributions are as follows:
% \vspace{-0.25cm}
\begin{itemize}
    \item We propose a generative video compression framework based on the Group-of-Latents (GoL) paradigm. By exploiting causal latent-space characteristics, our approach enables decoupled and flexible optimization of spatial and temporal quality under extreme bitrate constraints.
    
    \item We design the Deep Compression Module (I-DCM) to establish robust structural anchors. It employs entropy modeling to encode $I$-latents into ultra-compact representations, preserving essential perceptual textures with minimal bits.
    
    \item We introduce the Unified Latent Denoising Module (U-LDM) for unified temporal synthesis. This DiT-based module simultaneously refines $I$-latents and synthesizes $P$-latents from noise, reconstructing temporal dynamics at zero additional bitrate cost.
\end{itemize}

\vspace{-0.2cm}
% ==========================================================================
\section{Related Work}
\label{sec:relwork}

\subsection{Diffusion Model}
Diffusion models (DMs) have become the dominant paradigm in generative modeling, achieving state-of-the-art performance in both image and video synthesis. In the image domain, foundational models like Stable Diffusion~\cite{rombach2022high} pioneered denoising generation using U-Net architectures. Recently, the field has pivoted toward DiTs, exemplified by Flux~\cite{labs2025flux}, to exploit the scalability and expressive capacity of Transformer backbones. Parallel to this architectural shift, contemporary video generation models—such as CogVideoX~\cite{yang2024cogvideox}, Wan~\cite{wan2025wan}, and Cosmos~\cite{agarwal2025cosmos}—increasingly adopt DiT-based frameworks to model complex spatio-temporal dependencies. These advancements have significantly enhanced video fidelity and coherence, providing a robust foundation for downstream tasks such as controlled generation~\cite{ali2025world,wang2025language,geng2025motion,yuan2025vgdfr}.

\begin{figure*}[t]
  % \vskip 0.2in
  \begin{center}
    \centerline{\includegraphics[width=0.9\textwidth]{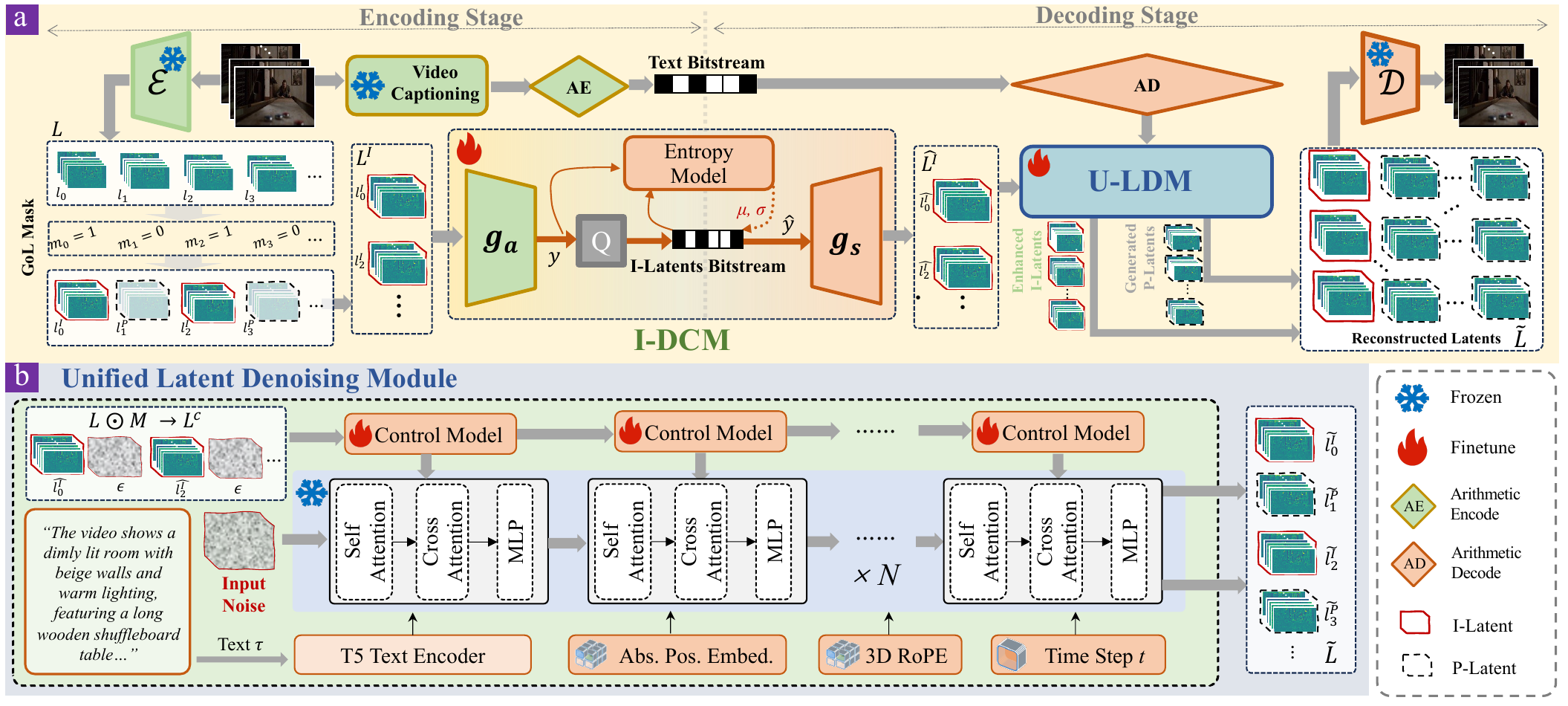}}
    \caption{Overview of the proposed framework. (a) Overall pipeline. The encoder compresses identified $I$-latents via I-DCM and extracts a text description into the bitstream. The decoder interleaves reconstructed $I$-latents with pure noise according to the GoL mask, and the U-LDM generates the complete latent sequence guided by this structural condition and the text prompt. \\ (b) Detailed architecture of the U-LDM.}
    \label{fig:overall_arch}
  \end{center}
  \vskip -0.2in
\end{figure*}

\subsection{Generative Image Compression}
Driven by advancements in deep representation learning, end-to-end image compression~\cite{minnen2018joint,he2022elic,li2024frequency} has steadily outpaced traditional codecs such as JPEG and BPG~\cite{wallace1992jpeg,bpg}. Foundational VAE-based architectures primarily focused on minimizing objective distortion in the pixel space~\cite{jiang2023mlic,xu2024idempotence}. To further enhance perceptual fidelity, subsequent research integrated adversarial discriminators from GANs~\cite{muckley2023improving,korber2025egic}, effectively mitigating blurring artifacts and improving subjective quality.

Diffusion models have further redefined the field by serving as powerful generative decoders~\cite{ma2024correcting,yang2023lossy,relic2024lossy}. While DMs generally improve subjective quality through multimodal priors~\cite{lei2023text,bachard2024coclico,kuang2024consistency}, they offer a pivotal advantage in extreme bitrate regimes. 
In these scenarios, their extensive generative priors facilitate the reconstruction of high-fidelity details from minimal codewords, effectively surmounting the information scarcity caused by severe quantization~\cite{careil2024towards,li2025toward,li2025rdeic,xu2025decouple,xue2025one}. While these generative techniques have matured in the image domain, extending them to model complex spatio-temporal dynamics in video compression remains an under-explored challenge.

\subsection{Video Compression}
\label{sec:relwork_video_compression}
Traditional codecs~\cite{sullivan2012overview,bross2021overview} have established a mature framework for video compression through handcrafted spatial prediction and motion-compensated residual coding. Building upon this foundation, VAE-based end-to-end frameworks~\cite{lu2019dvc,li2021deep,li2023neural,li2024neural,jia2025towards} replace handcrafted modules with learned transforms, with the DCVC series further pioneering implicit inter-frame modeling via latent-domain temporal propagation. Subsequent variants such as PLVC~\cite{yang2022perceptual} incorporated GAN-based discriminators~\cite{mentzer2022neural,salehkalaibar2024perception} to improve subjective quality, and GLC-Video~\cite{qi2025generative} further pushes VAE-based compression toward lower bitrates via VQ-VAE latent coding. Despite these advances, VAE-based architectures are fundamentally constrained by their limited model capacity and the absence of rich generative priors, rendering them unable to maintain acceptable perceptual quality under extreme bitrate constraints.

The emergence of large-scale diffusion models offers a promising avenue to overcome these limitations. Recent efforts can be broadly categorized into two paradigms. The first applies diffusion models as perceptual enhancers within conventional coding pipelines. DiffVC~\cite{ma2025diffusion}, DiffVC-OSD~\cite{ma2025diffvc}, and YODA~\cite{li2026yoda} refine decoded frames through pre-trained diffusion priors with varying degrees of temporal awareness. However, the underlying codec still carries the full bitstream burden, confining these methods to standard bitrate ranges, and their frame-level denoising lacks the joint spatio-temporal modeling needed for consistent inter-frame quality.

The second paradigm natively integrates diffusion models as the core reconstruction engine. Li et al.~\cite{li2024extreme} and Yi et al.~\cite{yi2025conditional} employ ControlNet-based adapters conditioned on compressed keyframes, T-GVC~\cite{wang2025tgvc} introduces sparse trajectory guidance to improve motion fidelity, and GNVC-VD~\cite{mao2025generative} performs sequence-level latent refinement by jointly denoising all compressed spatio-temporal latents through a video DiT. While these methods have advanced perceptual quality, they generally rely on conventional redundancy removal strategies without explicitly structuring temporal decomposition within the latent space, limiting their achievable bitrate floor. In contrast, our GoL framework performs explicit $I$/$P$ decomposition directly within the latent space of the causal tokenizer, natively aligned with the DiT architecture. By transmitting only structural anchors and synthesizing the remaining temporal components through a unified generative process, this formulation achieves more thorough redundancy elimination, pushing the achievable bitrate into a further extreme regime while maintaining flexible rate-quality trade-offs.

% ==========================================================================
\section{Method}
\label{sec:method}

\subsection{Overview}
As shown in Figure~\ref{fig:overall_arch}~(a), for a given video $X$, the pipeline begins by mapping it into the latent space $L$ via a pre-trained causal tokenizer (Sec.~\ref{sec32}). Following the proposed GoL configuration, this latent representation is partitioned into Intra-latents ($L^I$) and Inter-latents ($L^P$). At the encoding stage, the I-DCM Encoder (Sec.~\ref{sec33}) compresses $L^I$ into ultra-compact representations, while a semantic text description $\tau$ is extracted and encoded into the bitstream with negligible overhead. Together, the compressed $L^I$ and $\tau$ constitute the complete transmitted bitstream. At the decoding stage, the I-DCM Decoder first recovers the structural anchors $\hat{L}^I$ from the bitstream. The Unified Latent Denoising Module (U-LDM) (Sec.~\ref{sec34}) then jointly refines quantization artifacts in $\hat{L}^I$ and synthesizes the $L^P$ components from pure noise through a single generative process, guided by the text prompt $\tau$. This yields a complete, enhanced latent $\hat{L}$ without incurring any additional bitrate for the inter-frame components. Finally, $\hat{L}$ is projected back to the pixel domain by the tokenizer decoder to produce the reconstructed video $\tilde{X}$. The end-to-end training procedure is detailed in Sec.~\ref{sec35}.
\vspace{-0.2cm}

\subsection{Latent Space Mapping}
\label{sec32}

\begin{figure}[t]
  % \vskip 0.2in
  \begin{center}
    \centerline{\includegraphics[width=0.87\columnwidth]{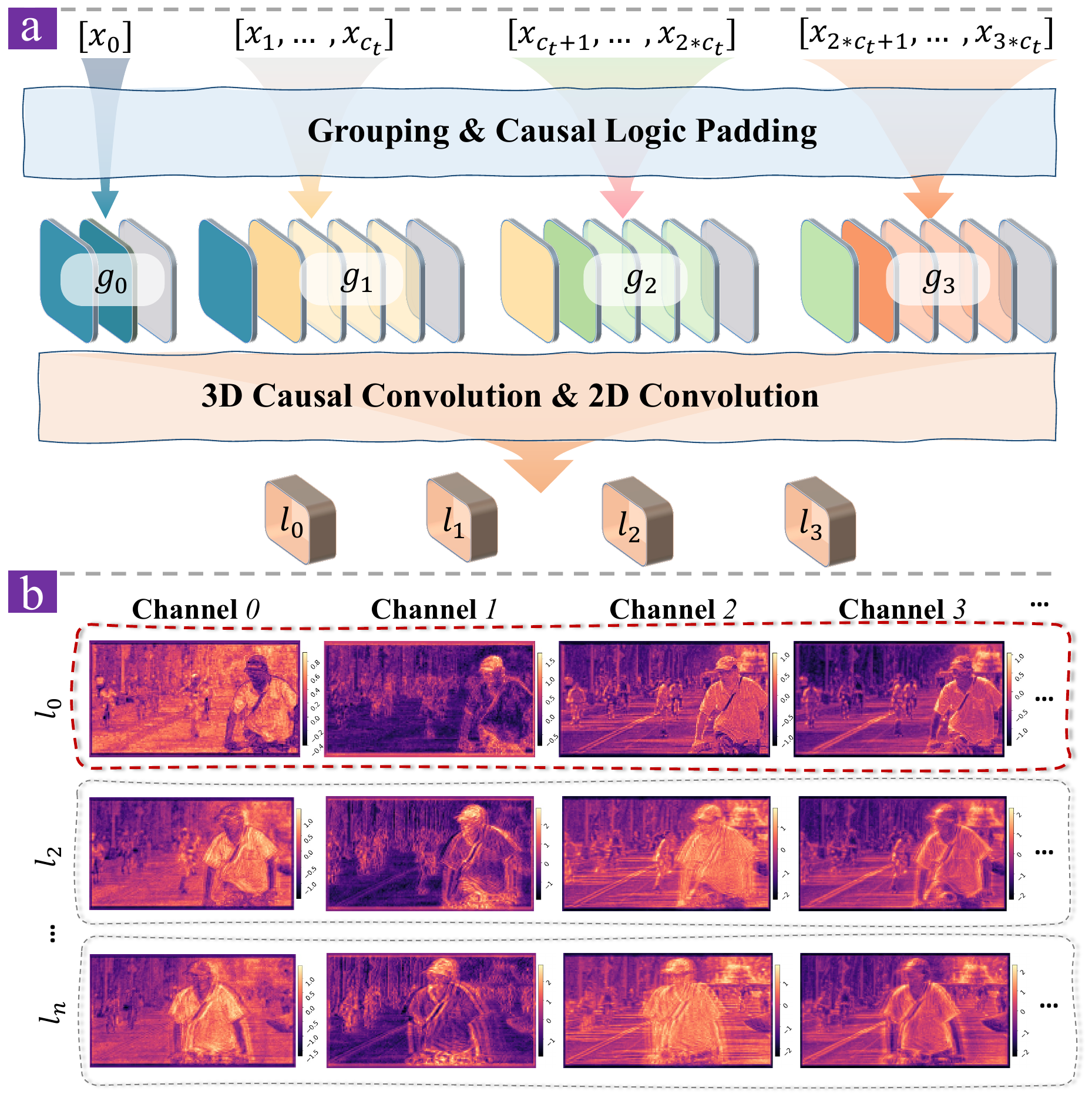}}
    \caption{The (a) encoding process of the causal tokenizer and (b) visualization of the latents.} 
    \label{fig:tokenize}
  \end{center}
  \vskip -0.1in
\end{figure}

Following the paradigm of modern video generation models, we employ a pre-trained tokenizer encoder $\mathcal{E}(\cdot)$~\cite{agarwal2025cosmos,wan2025wan} to map pixel-domain video $X = [x_0, \dots, x_{T-1}] \in \mathbb{R}^{T \times 3 \times H \times W}$ into a spatiotemporally downsampled latent space $L$ with dimensions $(\frac{T}{c_t} + 1) \times C \times \frac{H}{c_s} \times \frac{W}{c_s}$, where $c_t$ and $c_s$ denote the temporal and spatial downsampling factors, respectively. Unlike static image encoding, the tokenizer employs causal 3D convolutions to process the temporal dimension. As illustrated in Figure~\ref{fig:tokenize}~(a), frames are organized into groups $G = [g_0, g_1, \dots, g_{\frac{T}{c_t}}]$, where the initial group $g_0$ contains a single frame and subsequent groups each comprise $c_t$ frames. These groups are then transformed into a sequence of temporal latents $L = [l_0, l_1, \dots, l_{\frac{T}{c_t}}]$. The causal structure enables rich information interaction within each frame group for preserving spatial textures, while enforcing sequential dependency across groups to maintain temporal progression.

Visualization of the latent space in Figure~\ref{fig:tokenize}~(b) reveals a natural decomposition of spatial and temporal information. Each individual latent $l_i$ primarily encapsulates spatial textural features, while motion information is encoded through the transitions between consecutive latents $l_i$ and $l_{i+1}$. This observation motivates our proposed Group-of-Latents~(GoL), an explicit latent-domain counterpart of the Group-of-Pictures~(GoP) in traditional coding. GoL defines a temporal configuration through a binary mask:
\begin{equation}
    M = \{m_0, m_1, \dots, m_{\frac{T}{c_t}}\}, \quad m_i \in \{0, 1\},
    \label{eq:gol}
\end{equation}
where the value of each element $m_i$ determines the compression and reconstruction strategy for the corresponding latent position. Specifically, positions where $m_i = 1$ designate $I$-latents ($L^I \in \mathbb{R}^{C \times \frac{H}{c_s} \times \frac{W}{c_s}}$), which serve as structural anchors and are compressed to an extremely low bitrate via the I-DCM. Positions where $m_i = 0$ denote $P$-latents ($L^P \in \mathbb{R}^{C \times \frac{H}{c_s} \times \frac{W}{c_s}}$), which carry no transmitted information and are instead synthesized by the U-LDM from pure Gaussian noise at the decoder. By leveraging the generative capacity of the pre-trained DiT, the GoL framework enables flexible temporal configurations that decouple spatial fidelity from temporal reconstruction within a unified generative pipeline.

\subsection{I-latent Deep Compression}
\label{sec33}

The I-DCM is responsible for compressing the Intra-latents $L^I = \{L_i \mid m_i=1\}$ identified by the GoL mask $M$ into ultra-compact bitstreams. It adopts a hierarchical VAE architecture with an encoder-decoder structure, where the encoder maps $L^I$ into a minimal set of codewords and the decoder reconstructs the structural anchors from the transmitted bitstream.

\subsubsection{I-DCM Encoder}
\label{sec331}
To achieve extreme bitrate reduction, the analysis transforms ($g_a$, $h_a$) perform aggressive spatial downsampling to obtain the compact feature $y$ and hyper-prior $z$, which are then discretized via quantization $Q(\cdot)$:
\begin{align}
    y &= g_a(L^I) \downarrow_{8\times}, \quad z = h_a(y) \downarrow_{64\times}, \nonumber \\
    \bar{y} &= Q(y), \quad \bar{z} = Q(z).
    \label{eq:idcm_encoding}
\end{align}
The hyper-prior $\bar{z}$ parameterizes a conditional Gaussian distribution $p(\bar{y}|\bar{z})$, capturing spatial dependencies to minimize the entropy of $\bar{y}$. Combined with the spatio-temporal stride $c_t \times c_s \times c_s$ already introduced by the tokenizer $\mathcal{E}(\cdot)$, the additional $8\times$ and $64\times$ downsampling on $y$ and $z$ dramatically reduces the symbol space volume, minimizing the total codeword consumption after entropy coding.

In addition to the latent bitstream, we transmit a compact text description $\tau$ that captures the global semantic content of the video. This text prompt is losslessly compressed via standard zlib encoding and appended to the bitstream, contributing approximately $3 \times 10^{-5}$ bpp and imposing virtually no burden on the overall compression system.

\subsubsection{I-DCM Decoder}
\label{sec332}
At the decoding stage, the I-DCM Decoder recovers the quantized features $\bar{y}$ and $\bar{z}$ from the bitstream via arithmetic decoding, along with the text description $\tau$ via zlib decoding. The synthesis transforms ($g_s$, $h_s$) then reconstruct the Intra-latents as $\hat{L}^I = g_s(\bar{y} \mid h_s(\bar{z}))$, utilizing the spatial dependencies predicted by the hyper-prior network. The total bitrate of the transmitted bitstream is:
\begin{equation}
    \mathcal{R} = \mathbb{E} \left[ -\log_2 p_{\bar{z}}(\bar{z}) - \log_2 p_{\bar{y}|\bar{z}}(\bar{y} | \bar{z}) \right].
    \label{eq:idcm_rate}
\end{equation}
The detailed architecture of the I-DCM is provided in the Supplementary Material. While this process effectively establishes the structural anchors at an extremely low bitrate, the VAE-based reconstruction inherently prioritizes objective fidelity and lacks the capacity to recover perceptually rich textures. Moreover, the Inter-latents $L^P$ corresponding to masked positions carry no transmitted information and remain to be reconstructed. These two requirements, together with the recovered text description $\tau$ as semantic conditioning, jointly motivate the design of the U-LDM, which performs unified perceptual refinement and temporal synthesis as described in the following section.

\subsection{Unified Latent Refinement}
\label{sec34}
The U-LDM reconstructs the full video sequence by leveraging all information recovered from the compressed bitstream, including the quantized Intra-latents $\hat{L}^I$ from the I-DCM Decoder, the GoL configuration mask $M$ that defines the temporal structure, and the textual description $\tau$.

\subsubsection{Condition Assembly and Flow Matching Formulation}
\label{sec341}

To initiate the unified generative process, we construct a composite condition latent $L^c$ by interleaving the decoded structural anchors with pure Gaussian noise according to the GoL mask. Formally, $L^c$ is defined using the indicator function as:
\begin{equation}
    L^c = \mathds{1}_{m_i = 1} \hat{L}^I_i + \mathds{1}_{m_i = 0} \epsilon,
    \label{eq:latent_assembly}
\end{equation}
where $\mathds{1}_{\{\cdot\}}$ equals 1 when the condition meets the mask configuration and 0 otherwise, $\epsilon \sim \mathcal{N}(0, \mathbf{I})$ is pure Gaussian noise. This operation splices the spatial features of $\hat{L}^I$ with stochastic noise into a unified conditioning signal, where $I$-latent positions carry structural priors and $P$-latent positions are left entirely to generative synthesis.

The U-LDM is implemented as a DiT under the flow-matching framework, which predicts the velocity field to iteratively reconstruct clean data from noise as shown in Figure~\ref{fig:overall_arch}~(b). We formally define the underlying dynamics below.

\textbf{Definition 3.1} (\textit{Flow matching dynamics.})
\begin{equation}
    L_t = (1-t)\times L + t\times \epsilon.
    \label{eq:add_noise}
\end{equation}
Here, $L$ is a clean video latent and $\epsilon \sim \mathcal{N}(0, \mathbf{I})$ is ambient noise. The interpolated state $L_t$ at timestep $t \in [0, 1]$ is defined by the optimal transport linear interpolant.

\textbf{Definition 3.2} (\textit{Velocity field.})
\begin{equation}
    v_t = \frac{d L_t}{dt} = \epsilon - L.
    \label{eq:reverse}
\end{equation}
The ground-truth velocity $v_t$ is given by the time derivative of the flow path. Compared to stochastic diffusion processes, this velocity-based formulation yields straighter generation trajectories, resulting in improved training stability and higher sample quality for high-dimensional video data.

\subsubsection{Structural Conditioning and Velocity Prediction}
\label{sec342}

Unlike video generation tasks that prioritize creative diversity, video compression demands perceptual consistency, where the decoded output should faithfully reflect the original video content. This requirement implies that the generative process must be effectively constrained by the structural semantics preserved in the compressed bitstream.

A straightforward approach would directly concatenate $L^c$ with the noisy state, defining the input as $Concat(L_t, L^c)$. However, since $L^c$ contains non-Gaussian quantization artifacts from the I-DCM, this superposition corrupts the standard Gaussian marginals of $L_t$ and forces the model to learn a complex trajectory that deviates from the clean flow dynamics in Definition 3.1. To preserve the integrity of the flow matching process, we decouple the structural condition from the noisy state, allowing $L^c$ to act as a stationary external guidance rather than a component of the time-dependent input. Under this formulation, the velocity field prediction is decomposed as:
\begin{equation} 
    v_\theta(L_t, t \mid L^c) \approx v_{\text{fm}}(L_t, t) + \mathcal{H}(L^c),
    \label{eq:assumption_decomposition} 
\end{equation}
where $v_{\text{fm}}$ handles the Gaussian denoising dynamics defined in Equation~\ref{eq:add_noise}, and $\mathcal{H}$ represents a time-invariant structural bias derived from the GoL condition. Experimental validation of this design choice is provided in the Supplementary Material.

To implement this decoupled injection, we employ a ControlNet-based architecture as shown in Figure~\ref{fig:overall_arch}~(b), which injects $L^c$ into the deep layers of the DiT via zero-convolutions:
\begin{equation}
    \mathbf{f}_{out} = \mathcal{F}(\mathbf{f}_{in}, t, \tau) + \mathcal{Z}(\mathcal{G}_{copy}(L^c, t)),
    \label{eq:controlnet_injection}
\end{equation}
where $\mathcal{F}$ represents the main DiT block processing the pure noisy features $\mathbf{f}_{in}$, and $\mathcal{Z}$ denotes the zero-convolution layer. Crucially, the prediction target $v_t = \epsilon - L$ is derived from the original, uncompressed, and spatially complete latent $L$, which contains the full temporal sequence including both $I$- and $P$-latent positions. This ensures that the main branch learns to reconstruct the entire video latent, simultaneously refining quantization artifacts at $I$-latent positions and synthesizing missing content at $P$-latent positions, while the ControlNet branch provides stable structural guidance without corrupting the flow dynamics.

The U-LDM is trained by minimizing the velocity matching error against this complete target:
\begin{equation}
    \mathcal{L}_v = \mathbb{E}_{t, L, \epsilon} \left[ | v_\theta(L_t, t, \tau, L^c) - v_t |^2 \right].
    \label{eq:final_loss}
\end{equation}
Minimizing Equation~\ref{eq:final_loss} forces the U-LDM to integrate explicit supervision from the pristine target with the generative priors of the pre-trained DiT, producing the refined latent $\tilde{L}$ that is subsequently decoded by the tokenizer to yield the reconstructed video $\tilde{X}$. Furthermore, since the ControlNet branch provides rich and stable structural conditioning throughout the denoising trajectory, the model exhibits reduced dependence on extensive iterative sampling. As validated in our experiments, this property enables a significant reduction in denoising steps while maintaining high perceptual quality, offering a practical pathway toward lower decoding latency.

\begin{figure*}[t]
    \begin{center}
        \centerline{\includegraphics[width=0.9\textwidth]{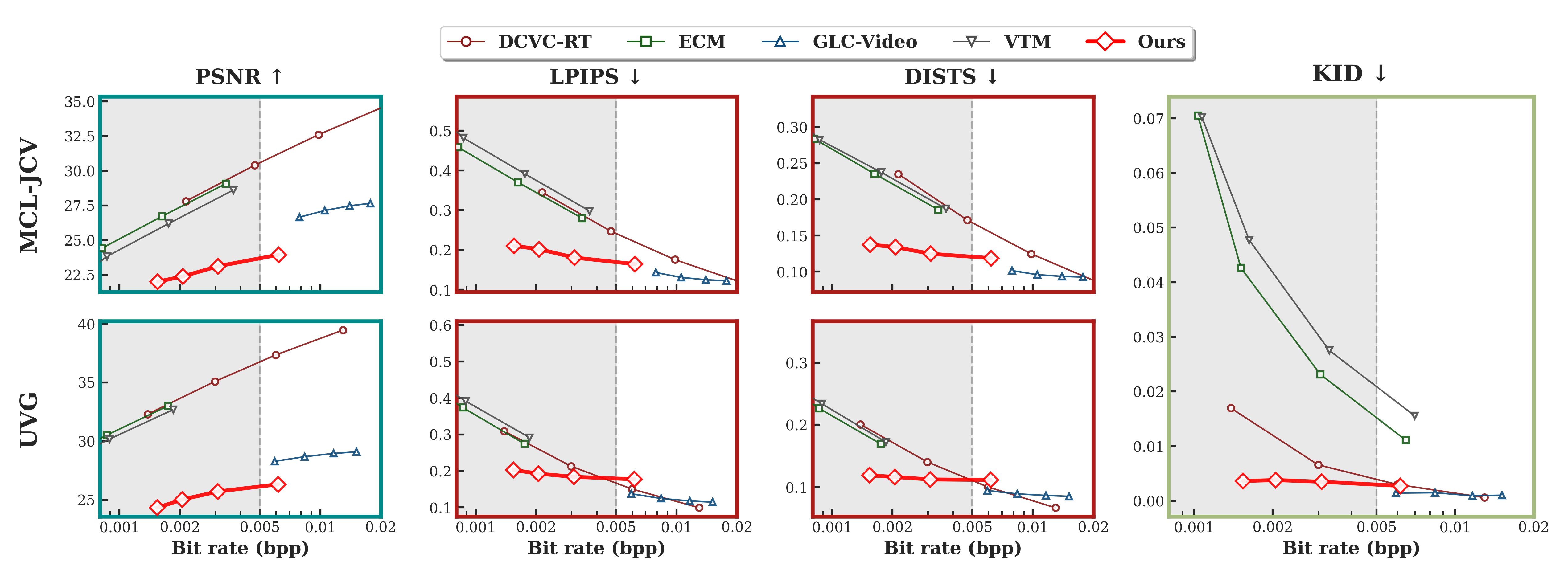}}
        \caption{Rate-distortion and rate-perception curve comparisons of different methods on the MCL-JCV and UVG. Bpp values are plotted on a logarithmic scale to facilitate clear comparisons within the extreme-low bitrate regime. The shaded area ($< 0.005$ bpp) represents the defined range for extreme-low bit rates. KID is calculated on the combined set of two datasets to ensure stable results.}
        \label{fig:rdrp}
    \end{center}
    \vskip -0.3in
\end{figure*}

\subsection{Training Recipe}
\label{sec35}

To achieve stable generative compression quality at extremely low bitrates, we propose a two-stage training recipe that progressively aligns the latent rate-distortion trade-off with the generative flow dynamics.

\textbf{Stage I: I-DCM Pre-training.} In the first stage, we train the I-DCM in isolation while keeping both the tokenizer and U-LDM frozen. The objective optimizes a rate-distortion trade-off in the latent space:
\begin{align}
    \mathcal{L}_{\text{stageI}} &= \lambda_{\text{I-DCM}} \underbrace{\text{MSE}(L^I, \hat{L}^I)}_{\text{Distortion}} + \underbrace{\mathcal{L}_{\text{R}}(\hat{y},\hat{z})}_{\text{Bitrate}},
    \label{eq:stage1_loss}
\end{align}
where the distortion term constrains the reconstruction error of the compressed Intra-latents, $\mathcal{L}_{\text{R}}$ estimates the bitstream volume via entropy modeling, and $\lambda_{\text{I-DCM}}$ adjusts the rate-distortion balance to favor optimization toward extremely low bitrates.

\textbf{Stage II: Joint Fine-tuning with U-LDM.} Building upon the completion of I-DCM alignment, the second stage jointly fine-tunes the I-DCM and U-LDM to train the denoising reconstruction capabilities:
\begin{align}
    \mathcal{L}_{\text{stageII}} &= \mathcal{L}_{\text{R}}(\hat{y},\hat{z}) + \mathcal{L}_v + \lambda_{\text{adp}} \{ \underbrace{\text{LPIPS}(X, \tilde{X})}_{\text{Perceptual}} + \underbrace{\text{MSE}(X, \tilde{X})}_{\text{Distortion}}\},
    \label{eq:stage2_loss}
\end{align}
where $\mathcal{L}_{\text{R}}$ maintains the bitrate constraint and $\mathcal{L}_v$ follows Equation~\ref{eq:final_loss}. Note that the latent-domain distortion term from Stage I is removed, as the velocity prediction objective $\mathcal{L}_v$ is mathematically equivalent to latent-level MSE supervision.

To directly supervise the quality of the reconstructed video in the pixel domain, we apply the LPIPS and MSE losses to a one-step projection estimate $\tilde{X}_{0|t} = \mathcal{D}(L_t - t \cdot v_\theta)$, which maps the current noisy state back to the data manifold. The derivation and implementation details of this projection are provided in the Supplementary Material. To regulate this pixel-level supervision, we employ a quadratic time-dependent annealing strategy defined by $\lambda(t) = \gamma (1-t)^2$. This non-linear weighting aggressively suppresses gradients during the chaotic early denoising phase ($t \approx 1$) while smoothly intensifying guidance as the trajectory converges ($t \to 0$), ensuring that pixel-level optimization is strictly contingent on the emergence of valid structural formations.

\textbf{Dynamic GoL Training.} Within the second stage, to ensure the U-LDM is robust across varying compression ratios, we employ a Dynamic GoL Training strategy. During each training iteration, the condition $L^c = L \odot M$ is synthesized using a stochastically generated binary mask $M = \{m_0, m_1, \dots, m_{T-1}\}$. We anchor the sequence with a mandatory Intra-latent at the first position ($m_0=1$), while randomly masking subsequent latents ($m_{i>0}$) with a variable dropout rate $p \in [0, 1]$:
\begin{equation}
    m_i \sim \text{Bernoulli}(1-p) \quad \text{for } i > 0.
\end{equation}
Since the bitrate is exclusively determined by the number of preserved Intra-latents ($m_i=1$), this stochastic training paradigm empowers a single U-LDM to generalize across diverse GoL configurations, supporting flexible bitrate adaptation during inference by simply adjusting the GoL mask without model retraining.

\vspace{-0.25cm}
% ==========================================================================
\section{Experiments}
\label{sec:exp}

\subsection{Implementation Details}
The causal tokenizer is adopted from Cosmos~\cite{agarwal2025cosmos}, with a temporal stride of $c_t=4$ and a spatial stride of $c_s=8$. The I-DCM follows an end-to-end VAE architecture, and the U-LDM is initialized from the pre-trained 2B-parameter DiT of Cosmos-Transfer2.5~\cite{ali2025world}. As discussed in Sec.~\ref{sec342}, the stable structural conditioning provided by our ControlNet-based injection reduces the model's dependence on extensive iterative sampling, enabling us to lower the denoising steps from the default 35 to 5 while maintaining high perceptual quality. To enforce extreme compression, the rate-distortion trade-off parameter is set to $\lambda_{\text{I-DCM}} = 10^{-3}$ in Equation~\ref{eq:stage1_loss}. All experiments were conducted with 8 NVIDIA A800 (80GB) GPUs. The discussion on the choice of the base DiT model, comprehensive architectural specifications, and hyperparameter settings are provided in the Supplementary Material.

\subsection{Experimental Setup}
\textbf{Datasets.}
We train on VidGen~\cite{tan2024vidgen} and evaluate on JVET-CTC (Class B and E)~\cite{ctc_test}, UVG~\cite{mercat2020uvg}, and MCL-JCV~\cite{wang2016mcl}. All video sequences are standardized to $1280 \times 704$ resolution to align with the native resolution of the pre-trained DiT. The number of frames is set to 93. For text descriptions, the VidGen training set natively provides high-quality captions, while we employ Gemini 3~\cite{gemini} to annotate the test set sequences.

\textbf{Metrics.}
Compression efficiency is measured in Bits-Per-Pixel (bpp) within the extreme-low bitrate range ($<0.005$ bpp). While we report PSNR and Structural Similarity (SSIM) for reference, pixel-wise fidelity metrics are known to diverge from human perception at extreme-low bitrates. We therefore adopt LPIPS~\cite{zhang2018unreasonable} and DISTS~\cite{ding2020image} as primary metrics for perceptual quality assessment, and Kernel Inception Distance (KID)~\cite{binkowski2018demystifying} for evaluating distribution-level realism. A detailed discussion on the rationale and properties of these metrics is provided in the Supplementary Material.

\textbf{Comparison methods.}
We benchmark against the latest standard VVC (VTM)~\cite{vtm} and the next-generation fusion framework (ECM)~\cite{ecm}. For neural compressors, we select DCVC-RT~\cite{jia2025towards} as the state-of-the-art rate-distortion optimized model and GLC-Video~\cite{qi2025generative} as the state-of-the-art generative low-bitrate codec. Perceptual methods such as PLVC~\cite{yang2022perceptual} are omitted as they fail to reach the target bitrate ($< 0.005$ bpp). For concurrent diffusion-based video codecs discussed in Sec.~\ref{sec:relwork_video_compression}, due to differences in target bitrate ranges, experimental configurations, and the absence of publicly available implementations, we provide a qualitative discussion based on reported results. Detailed configurations are provided in the Supplementary Material.

\subsection{GoL Configuration and Rate Points}
\label{sec_gol_config}
A distinctive advantage of our framework is that a single trained model supports multiple bitrate points through GoL mask configuration alone, without any model retraining. Our evaluation begins with the ``All-Intra'' mode using an all-ones mask $[1111]$, where every latent is treated as an $I$-latent and compressed via I-DCM, achieving approximately $0.006$ bpp after U-LDM refinement. Subsequent rate points are obtained by varying the GoL mask to reduce the proportion of transmitted $I$-latents. Specifically, the $[1010]$, $[1001]$, and $[1000]$ configurations retain every second, every third, and every fourth latent as $I$-latents, reducing the bitrate to approximately $1/2$, $1/3$, and $1/4$ of the All-Intra baseline, respectively. While the model employs stochastic Bernoulli masking during training (Sec.~\ref{sec35}) for robustness, we adopt these evenly distributed patterns during testing to ensure evaluation stability. As shown in Figure~\ref{fig:rdrp}, the four resulting rate points form a smooth and continuous rate-perception curve, validating that the Dynamic GoL Training strategy effectively generalizes across diverse mask configurations unseen during training.

\begin{table}[t]
    \centering
    \footnotesize
    \renewcommand{\arraystretch}{1.0}
    \setlength{\tabcolsep}{4pt} 
    \setlength{\aboverulesep}{0pt}
    \setlength{\belowrulesep}{0pt}
    \caption{Model-based perceptual quality evaluation results.}
    \label{tab_vlm_eval}
    \begin{tabularx}{\columnwidth}{c y{80} x{50} >{\centering\arraybackslash}X}
        \toprule
        {\scshape Evaluator} & {\scshape Dimension} & {\scshape GLC-Video} & {\scshape Ours}  \\ 
        \midrule
        
        \multirow{4}{*}{Opus 4.6} 
        & Texture Flickering$\uparrow$     & 4.87 & \textbf{6.23}  \\
        & Temporal Consistency$\uparrow$   & 5.56 & \textbf{6.57}  \\
        & Spatial Detail$\uparrow$         & 4.13 & \textbf{4.72}  \\
        & \cellcolor{softblue!40}Overall Quality$\uparrow$         & \cellcolor{softblue!40}4.67 & \cellcolor{softblue!40}\textbf{5.45}  \\
        \midrule
        
        \multirow{4}{*}{GPT 5.4} 
        & Texture Flickering$\uparrow$     & 6.30 & \textbf{7.10}  \\
        & Temporal Consistency$\uparrow$   & 5.40 & \textbf{6.30}  \\
        & Spatial Detail$\uparrow$         & 5.20 & \textbf{5.60}  \\
        & \cellcolor{softblue!40}Overall Quality$\uparrow$         & \cellcolor{softblue!40}5.63 & \cellcolor{softblue!40}\textbf{6.33}  \\
        \bottomrule
    \end{tabularx}
    \vspace{-0.1cm}
\end{table}

\begin{table}[t]
    \centering
    % \tiny
    \footnotesize
    \renewcommand{\arraystretch}{1.0}
    \setlength{\tabcolsep}{2pt} 
    \setlength{\aboverulesep}{0pt}
    \setlength{\belowrulesep}{0pt}
    \caption{Quality-efficiency trade-off across denoising steps. We report inference time and generation quality metrics. Test platform is a single A800.}
    \label{tab:dit_steps}
    \begin{tabularx}{\columnwidth}{l l x{25} x{25} x{26} x{26} x{26} >{\centering\arraybackslash}X}
        \toprule
        {\scshape Method} & {\scshape Steps} & {\scshape Enc. (s)} & {\scshape Dec. (s)} & {\scshape PSNR $\uparrow$} & {\scshape SSIM $\uparrow$} & {\scshape LPIPS $\downarrow$} & {\scshape DISTS $\downarrow$} \\ 
        \midrule
        VTM & -- & 9767.56 & \underline{18.83} & 28.60 & 0.79 & 0.30 & 0.18 \\
        ECM & -- & 11389.50 & 21.48 & \underline{29.07} & \underline{0.81} & 0.28 & 0.19 \\
        DCVC-RT & -- & \textbf{2.44} & \textbf{2.58} & \textbf{30.40} & \textbf{0.83} & 0.25 & 0.17 \\
        GLC-Video & -- & 29.09 & 48.91 & 26.66 & 0.74 & \textbf{0.14} & \underline{0.10} \\
        \midrule
        \multirow{4}{*}{Ours} & 1  & \underline{18.13} & 33.01   & 24.23 & 0.76 &  0.28 & 0.18 \\
        % \rowcolor{softblue!40}
                              & \cellcolor{softblue!40}5  & \cellcolor{softblue!40}\underline{18.13} & \cellcolor{softblue!40}130.23 & \cellcolor{softblue!40}24.01 & \cellcolor{softblue!40}0.71 & \cellcolor{softblue!40}0.16 & \cellcolor{softblue!40}0.11 \\
                              & 10 & \underline{18.13} & 234.01 & 23.71 & 0.67 &  0.16 & \textbf{0.09} \\
                              & 20 & \underline{18.13} & 469.41 & 23.50 & 0.66 &  \underline{0.15} & 0.12\\
        \midrule
        
        \bottomrule
    \end{tabularx}
    \vspace{-0.2cm}
\end{table}

\subsection{Comparison Results}
\label{sec_comparison}

The rate-distortion and rate-perception curves are presented in Figure~\ref{fig:rdrp}. Our method consistently operates within the extreme-low bitrate range ($< 0.005$ bpp) across all GoL configurations. While traditional codecs and DCVC-RT can technically reach these bitrates through aggressive quantization, they suffer from severe perceptual degradation. On the perceptual metrics LPIPS and DISTS, our method demonstrates a pronounced advantage, maintaining a stable trajectory as bitrate decreases, whereas baseline approaches degrade rapidly due to accumulating compression artifacts. Similarly, our method maintains near-zero KID scores across all rate points, indicating close alignment with the natural video distribution. Although traditional methods generally yield higher PSNR at comparable bitrates, pixel-wise fidelity metrics become unreliable in this regime, as a reconstruction can achieve statistically reasonable PSNR while exhibiting catastrophic visual quality such as pervasive blurring and loss of structural detail. This phenomenon is consistent with the theoretical predictions of RDP under extreme rate constraints.

\begin{table}[t]
    \centering
    \footnotesize
    \renewcommand{\arraystretch}{1.0}
    \setlength{\tabcolsep}{2pt} 
    \setlength{\aboverulesep}{0pt}
    \setlength{\belowrulesep}{0pt}
    \caption{Ablation experiments on U-LDM. The test dataset is MCL-JCV. The arrow direction indicates better quality.}
    \label{tab_abla_uldm}
    \begin{tabularx}{\columnwidth}{y{50} x{40} x{40} x{40} >{\centering\arraybackslash}X}
        \toprule
        {\scshape Method} & {\scshape PSNR$\uparrow$} & {\scshape SSIM$\uparrow$} & {\scshape LPIPS$\downarrow$} & {\scshape DISTS$\downarrow$}  \\ 
        \midrule
        
        w/o U-LDM & 19.93 & 0.64 & 0.43 & 0.29  \\
        \rowcolor{softblue!40}
        w/ U-LDM & 24.01 & 0.71 & 0.16 & 0.11  \\
        % \midrule
        
        \bottomrule
    \end{tabularx}
    \vspace{-0.1cm}
\end{table}

\begin{figure}[t]
  % \vskip 0.2in
  \begin{center}
    \centerline{\includegraphics[width=0.9\columnwidth]{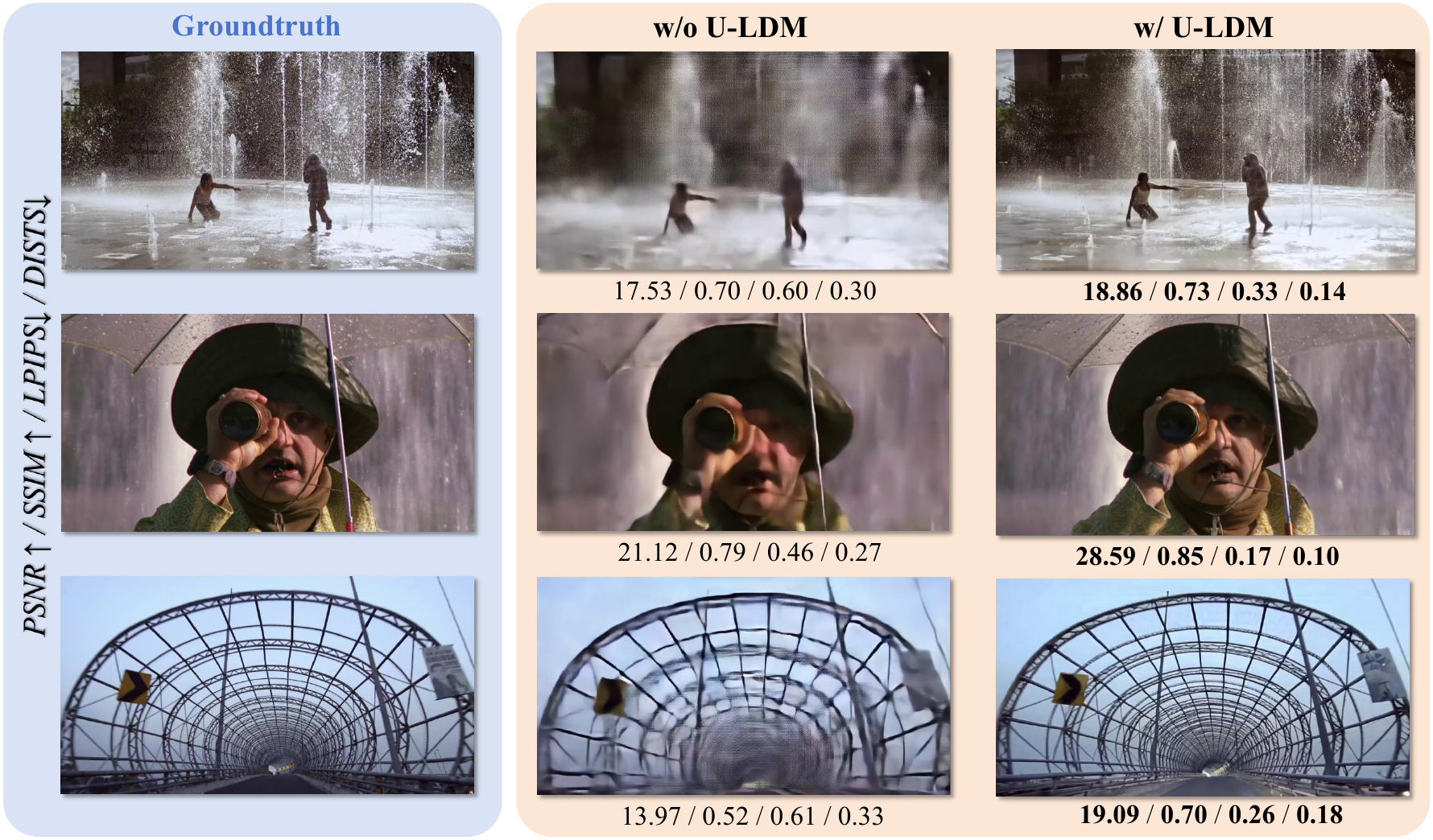}}
    \caption{Visualization results and metric comparisons of ablation experiments on U-LDM.} 
    \label{fig:abla}
  \end{center}
  \vskip -0.3in
\end{figure}

Among generative approaches, GLC-Video exhibits competitive perceptual quality at moderate low bitrates, but our method extends high-fidelity generation further into the extreme-low regime, validating that the explicit decoupling of spatial anchors from temporal synthesis enables realistic reconstruction with substantially less transmitted information.
As for concurrent diffusion-based video codecs, these methods generally operate at considerably higher bitrates~\cite{ma2025diffusion,wang2025tgvc,mao2025generative}. Our GoL formulation reaches a lower bitrate floor by synthesizing $P$-latents at zero additional cost, suggesting that explicit latent-domain temporal decomposition is a direction for further bitrate reduction.

\subsection{Model-based Perceptual Evaluation}
To evaluate perceptual quality at extreme bitrates, we employ the ``LLM-as-Judge" paradigm~\cite{liu2025your} using Opus~\cite{opus} and GPT~\cite{gpt}, which aligns closely with human perception in semantic and temporal dimensions. Through structured prompts, we assess generative artifacts by analyzing the consistency of object appearance, motion, and scene transitions~\cite{liu2024evalcrafter,liu2024survey}. As reported in Table~\ref{tab_vlm_eval}, our GoL framework significantly outperforms the generative baseline GLC-Video. The marked improvements in Temporal Consistency and Spatial Detail further validate our strategy’s ability to preserve structural stability and motion realism. 

\begin{figure*}[t]
    \begin{center}
        \centerline{\includegraphics[width=0.9\textwidth]{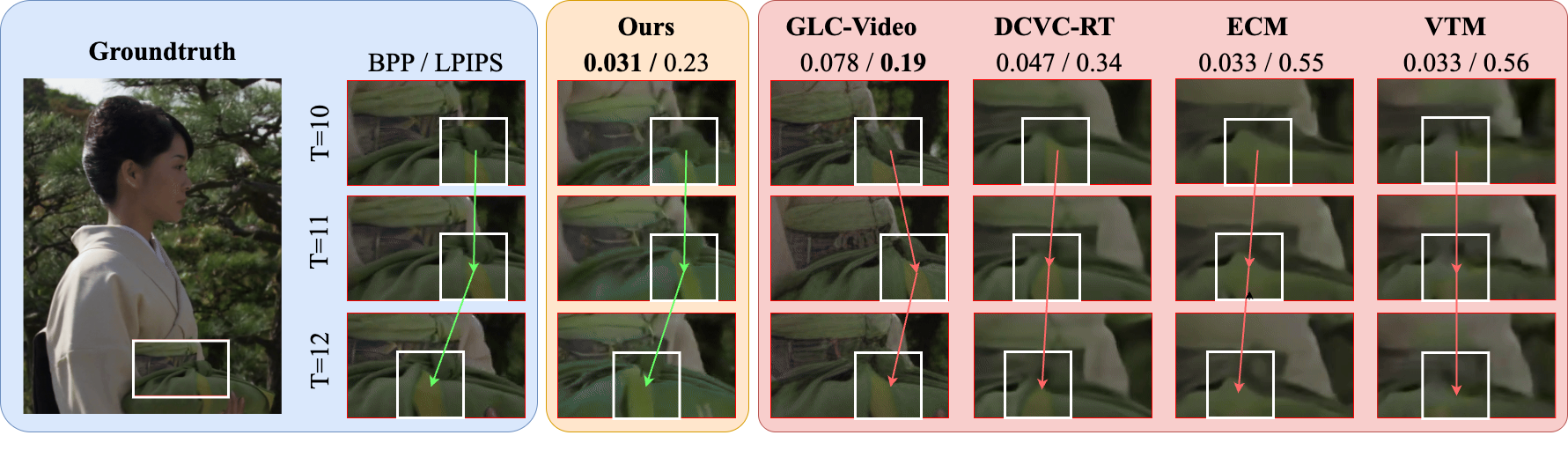}}
        \caption{Visual comparison at a comparable bitrate across all methods. This operating point is selected to enable fair cross-method evaluation and does not represent our lowest achievable bitrate.}
        \label{fig:visual}
    \end{center}
    \vskip -0.1in
\end{figure*}

\begin{figure}[t]
    \begin{center}
        \centerline{\includegraphics[width=0.92\linewidth]{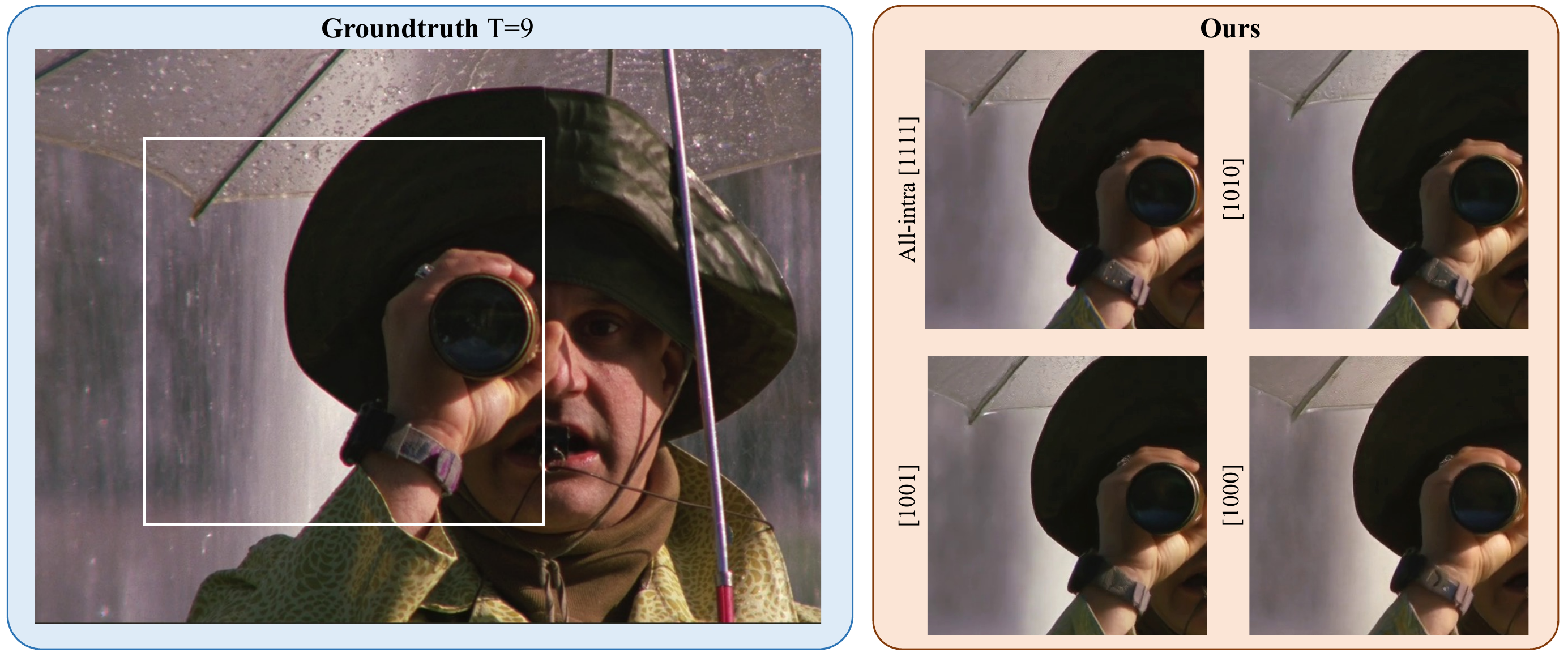}}
        \caption{Visual results of our method under different GoL configurations.}
        \label{fig:visual_gol}
    \end{center}
    \vskip -0.2in
\end{figure}

\subsection{Ablation Study}
We evaluate the contribution of U-LDM by comparing the full pipeline against a baseline that bypasses diffusion refinement under the All-Intra $[1111]$ mode. This configuration isolates the intra-frame refinement capability of U-LDM, as the GoL mask preserves all latent positions and no temporal synthesis is involved. The effectiveness of U-LDM on inter-frame synthesis has already been demonstrated through the GoL rate points in Sec.~\ref{sec_gol_config}, where $P$-latents are entirely generated by U-LDM at zero bitrate cost. As shown in Table~\ref{tab_abla_uldm}, U-LDM brings consistent gains across all metrics with the bitstream unchanged. Qualitatively, Figure~\ref{fig:abla} confirms that while I-DCM preserves structural anchors, U-LDM is indispensable for recovering fine details and synthesizing realistic textures. Additional ablation studies on text conditioning and other design choices are provided in the Supplementary Material.

\subsection{Denoising Steps and Complexity}
\label{sec_complexity}
Table~\ref{tab:dit_steps} details the computational complexity and performance trade-offs across different denoising steps. As discussed in Sec.~\ref{sec342}, the stable structural conditioning provided by our ControlNet-based injection enables effective reconstruction with significantly fewer denoising steps. Our default configuration adopts 5 steps, achieving a favorable balance between perceptual quality and decoding efficiency. At this setting, encoding is fast due to the lightweight I-DCM architecture, and decoding latency is comparable to existing efficiency-optimized diffusion-based video compressor~\cite{mao2025generative} while being substantially lower than methods employing native multi-step diffusion processes~\cite{wang2025tgvc,ma2025diffusion}. We further evaluate the effect of increasing denoising steps beyond our default 5-step setting at test time. As shown in Table~\ref{tab:dit_steps}, additional steps progressively improve perceptual metrics while decreasing objective fidelity, reflecting the inherent perceptual preference of the DiT prior and the distortion-perception antagonism established by RDP theory. Future work on diffusion distillation or native single-step generative models could further reduce decoding latency.

\begin{figure}[t]
  % \vskip 0.2in
  \begin{center}
    \centerline{\includegraphics[width=\columnwidth]{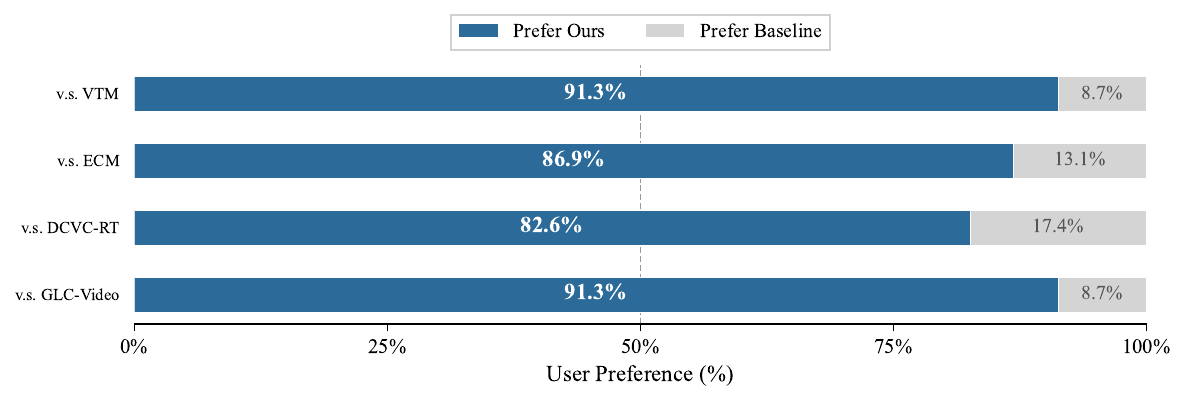}}
    \caption{User study result. The bars depict the preference rates for our method in pairwise comparisons.} 
    \label{fig:user}
  \end{center}
  \vskip -0.2in
\end{figure}

\subsection{Visual Results and User Study}
\label{sec_visual_user}
The visualizations in Figure~\ref{fig:visual} and Figure~\ref{fig:visual_gol} demonstrate our advantages in both spatial perception and temporal stability. In the cross-method comparison of Figure~\ref{fig:visual}, all baselines are evaluated near their lowest achievable bitrate, yet our method attains competitive perceptual quality at an even lower bitrate while maintaining stable object textures across the three-frame $I$/$P$ transition, as highlighted by the white boxes. In Figure~\ref{fig:visual_gol}, the selected frame falls into a $P$-latent position under the $[1001]$ and $[1000]$ configurations, meaning it is entirely synthesized without any transmitted information. The reconstructed quality remains largely consistent with the All-Intra baseline, with only subtle differences in background details, demonstrating the robustness of the $I$/$P$ decoupling.

To further assess subjective quality, we conducted a user study using a two-alternative forced-choice (2AFC) protocol, benchmarking against VTM, ECM, DCVC-RT, and GLC-Video, consistent with the codecs in our quantitative experiments. The detailed experimental setup is described in the Supplementary Material. As illustrated in Figure~\ref{fig:user}, our method received strong user preference across all pairwise comparisons, achieving over 85\% preference against both traditional and neural codecs. These subjective findings provide complementary validation of our advantages in spatial detail and temporal stability.

% ==========================================================================
\section{Conclusion}
\label{sec:conclusion}
Motivated by the goal of achieving extreme-low bitrates, we explore the minimal information required for effective video representation and propose a unified generative framework that prioritizes perceptual quality. Our approach explicitly decouples the preservation of spatial structure from the synthesis of temporal motion. Guided by the GoL strategy, the I-DCM minimizes transmission overhead by compressing only key structural anchors that retain essential spatial information. The U-LDM then exploits pre-trained DiT priors to perform unified reconstruction, leveraging conditional guidance to refine intra-frame textures and synthesize inter-frame motion dynamics with high perceptual realism. Through a meticulously designed training recipe, our framework operates at extreme-low bitrates ($< 0.005$ bpp), with experimental results demonstrating superior spatial detail and robust temporal consistency.

% \begin{acks}
% \end{acks}

\bibliographystyle{ACM-Reference-Format}
\bibliography{sample-base}

% \appendix









\end{document}